# Design Scheme for Mach-Zehnder Interferometric CWDM Wavelength Splitters/Combiners

**Matteo Cherchi[†]**

Pirelli Labs – Optical Innovation, viale Sarca 222, Milan, Italy

Abstract: We propose an analytical approach to design flattened wavelength splitters with cascaded Mach-Zehnder interferometers when wavelength dependence of the directional couplers cannot be neglected. We start from a geometrical representation of the action of a doubly point symmetrical filter, assuming no wavelength dependence of the couplers. Next we derive the analytical formulas behind its working principle and we extend them to the wavelength dependent case. We also show how the geometrical representation allows to broaden the class of working structures.

---

[†] He is now with Dipartimento di Ingegneria Elettrica, Università di Palermo, 90128 Palermo, Italy

# Introduction

Wavelength splitters/combiners are a particular kind of wavelength multi/demultiplexers, which are meant to manage two bands only. They can be designed and implemented in very different ways. In the context of planar lightwave circuits, a convenient choice is the cascading of Mach-Zehnder interferometers.

Standard techniques for the synthesis [1]-[3] of this kind of filters don't take into account the wavelength dependence of directional couplers. This means that they can be applied to DWDM circuits only, within bands where this dependence is negligible. On the other hand Jinguji et al. [4] proposed the point symmetrical configuration as an alternative method for filter synthesis. They applied these technique to the two limiting cases of strong wavelength dependence and no wavelength dependence of couplers. In the first case wavelength selection is due to the coupler response, whereas in the second case is due to the extra-length of one of the two arms of each single interferometer.

We propose in this paper an extension of the last configuration to the intermediate case in which the wavelength dependence of couplers is not negligible, but not so strong to effectively allow wavelength selection. This is a typical situation when dealing with CWDM filters. In the light of a geometrical interpretation of the results in Ref. [4], we will show how to correct this wavelength sensitivity to get still flat passbands. Even though the novel geometrical and analytical approach is presented in application to a practical case, it should be clear that what is proposed is not a particular structure but a new method to design and analyze interferometric filters. The presented application will show how the proposed method can give more physical insight and design control than any numerical approach.

# Wavelength independent couplers

In this section we will briefly present the synthesis technique based on doubly point-symmetrical cascading of Mach-Zehnder interferometers, supposing wavelength independent couplers. Then we will introduce the geometric representation of couplers and phase shifters that enables to reinterpret previous numerical results in terms of general analytic formulas.

## *The Doubly Point-Symmetrical Configuration*

Wavelength splitters/combiners are two-port devices that split/combine two bands centered at two different wavelengths $\lambda_1$ and $\lambda_2$.

The point-symmetric configuration has been introduced in Ref. [4] to get flat filtering with interferometric structures. In particular when repeated twice (see Fig. 1), this method ensures flat response both on cross-port and through port. In Ref. [4] the general working principle of this technique is explained analytically, whereas the lengths of the couplers are calculated using a numerical optimization algorithm. Couplers are supposed to be wavelength independent, whereas the extra-length of each Mach-Zehnder interferometer is chosen so that its phase shift is $n\pi$ ($n$ integer) at $\lambda_1$, and $(n+1)\pi$ at $\lambda_2$. Clearly, when $n$ is odd ($n+1$) is even and vice versa. For simplicity we will refer to even and odd phase shifts, depending on the parity of $n$.

We will now introduce a geometric representation for the action of an interferometer that not only makes it possible to analytically recover the results of the numerical approach, but also allows to broaden the class of working structures.

*Geometric Representation*

In two recent papers [5], [6] we have shown how a generalized Poincaré sphere representation [7]-[9] can be very helpful when studying wavelength dependent couplers and interferometers. The actions of both couplers and phase shifters can be represented as rotations on a spherical surface, analogous of the Poincaré sphere for polarization states. Figure 2 displays all the intersections of the $S_1$, $S_2$, and $S_3$ axes with the sphere. They represent respectively the single waveguides modes $E_1$, $E_2$ and their linear combinations $E_{S,A} \equiv \frac{1}{\sqrt{2}}(E_1 \pm E_2)$ and $E_{R,L} \equiv \frac{1}{\sqrt{2}}(E_1 \pm i E_2)$. Also the generic normalized state $P \equiv a_1 E_1 + a_2 E_2 \equiv \cos\alpha\, E_1 + \exp(i\theta)\sin\alpha\, E_2$ is plotted. Notice that the relative phase angle $\theta$ is exactly the altitude angle of the point P with respect to the equatorial plane on the circle perpendicular to the $S_1$ axis passing through P. Considering the cone having this very circle as basis and the sphere center as vertex, the power-splitting angle $\alpha$ is equal to half the half cone opening angle. Since we deal with lossless elements only, all the physical trajectories are limited to points on the spherical surface. Furthermore, when considering transformations which are compositions of reciprocal elements only (which is the case when dealing with directional couplers and phase shifters), all trajectories on the sphere can be compositions of rotations about axes belonging to the equatorial plane only [5], i.e. the $S_1 S_2$ plane. In particular the action of a phase shifter is represented by a rotation about the $S_1$ axis, and the action of a synchronous coupler is represented by a rotation about the $S_2$ axis. Hence, if we deal with synchronous couplers and $n\pi$ ($n$ integer) phase shifts only, we can restrict ourselves to the projection of the sphere on the $S_1 S_3$ plane, i.e. a 1-dimensional representation on the circle laying in that plane.

Let us define the angular expression of the amplitude coupling ratio [4], [5] of a coupler

$$\phi(\lambda) = \kappa(\lambda)[L + \delta L(\lambda)],  \qquad (1)$$

where $\kappa(\lambda)$ is the coupling per unit length in the straight part of the coupler and $\delta L(\lambda)$ accounts for the contribution of the input and output curves. We will assume these two quantities to be the same for all couplers, whereas we will let the coupler length $L$ vary from coupler to coupler. For the moment we will also assume any dependence of $\phi$ on $\lambda$ to be negligible.

Fig. 3 shows the projected trajectories on the $S_1 S_3$ plane of a Mach-Zehnder interferometer in the two cases of even and odd multiples of $\pi$ for the phase shift. $E_1$ and $E_2$ represent respectively the situation of all power in the input arm and all power in the other arm. At the light of what above noticed regarding the power splitting angle $\alpha$, the action of a coupler is simply represented by a rotation of twice the coupling angle $\phi$. A $2n\pi$ phase shift lives the system as it is whereas a $(2n+1)\pi$ phase shift sends any point of the circle to its mirror image with respect to the $S_1$ axis.

Hence, in the case of an even phase shift, the interferometer follows the path $E_1FG$, acting as a coupler with angle $\phi^A + \phi^B$, whereas in the case of an odd phase shift it will follow the path $E_1FGH$, equivalent to a coupling angle $\phi^B - \phi^A$.

## Analytic formulas

Keeping this in mind, we go back to the point-symmetric structure in Fig. 1. The filter works if the couplers satisfies the following conditions:

$$\begin{cases} 4\phi^A - 2\phi^B = t\dfrac{\pi}{2} + k\pi \\ 4\phi^A + 2\phi^B = (1-t)\dfrac{\pi}{2} + m\pi \end{cases}, \qquad (2)$$

where $t \in \{0,1\}$, $\phi^A$ and $\phi^B$ must be positive, $k$ is an integer and $m$ must be a non-negative integer. For $t = 0$ the first condition reads that for odd phase shifts power must remain in the through port and the second one means that for even phase shifts light must go in the cross port. For $t = 1$ the two ports exchange their role. Wavelength selectivity is simply obtained with a proper choice of the extra length $\Delta L \approx \pi / |\beta(\lambda_2) - \beta(\lambda_1)|$, being $\beta(\lambda)$ the propagation constant of the single waveguide mode. Solving (2) for $\phi^A$ and $\phi^B$, we get

$$\begin{cases} \phi^A = \dfrac{\pi}{16} + (m+k)\dfrac{\pi}{8} \\ \phi^B = \dfrac{\pi}{8} + (m-k-t)\dfrac{\pi}{4} \end{cases}, \quad (3)$$

where $m$ and $k$ obey the selection rules

$$\begin{cases} m \geq |k| & \text{for } t = 0 \\ m \geq k+1 & \text{for } t = 1 \text{ AND } k \geq 0 \\ m \geq -k & \text{for } t = 1 \text{ AND } k < 0 \end{cases}. \quad (4)$$

In particular for $t = 0$, $m = 2$, and $k = -1$ we find $\phi^A = 0.1875\,\pi$ and $\phi^B = 0.875\,\pi$ that are very close to the optimal values $\phi^A = 0.1882\,\pi$ and $\phi^B = 0.8626\,\pi$ numerically found in Ref. [4].

It is straightforward to rewrite (3) in terms of $L_X = \phi^X / \kappa - \delta L$ ($X = A, B$), clearly with the further constraint $\phi^X / \kappa > \delta L$.

Notice that the results of the numerical approach in Ref. [4] depend on the initial guess. On the contrary our approach allows to know all the solutions at once, and this may be useful, for example, to find the shortest one. In our formalism the choice $t = 0$, $m = 0$, $k = 0$ gives the smaller overall coupling angle $4\phi_i^A + 2\phi_i^B = \pi/2$, that is five times smaller than the coupling angle of the $t = 0$, $m = 2$, $k = -1$ case proposed in Ref. [4].

We like also to point out that, when the full Poincaré sphere representation is considered, it becomes clear why point-symmetry can guarantee band flatness, i.e. tolerance to wavelength changes, or, equivalently, insensitivity to changes of the phase shifter rotations. It will be shown in detail later on that the change of a given phase shift from its nominal value (due to a small deviation of the wavelength from $\lambda_i$) will be compensated for by a corresponding change in its point-symmetrical counterpart.

## Wavelength dependent couplers

In this section we will extend previous results to the case of wavelength dependent couplers. With the aid of the full Poincaré sphere representation we will also show that, for some choice of parameters *t*, *m* and *k*, the singly point-symmetrical configuration must be preferred in order to get well flattened response of the filter.

### *Generalized formulas*

Whenever it is required to manage two channels only, we can just focus on the spectral response within the two bands about $\lambda_1$ and $\lambda_2$. We can take into account the wavelength dependence rewriting (3) as

$$\begin{cases} 4\phi_2^A - 2\phi_2^B = t\dfrac{\pi}{2} + k\pi \\ 4\phi_1^A + 2\phi_1^B = (1-t)\dfrac{\pi}{2} + m\pi \end{cases} \quad (5)$$

where $\phi_i^X \equiv \kappa_i(L_X + \delta L_i) \equiv \kappa(\lambda_i)[L_X + \delta L(\lambda_i)]$, and we have assumed $\lambda_1$ ($\lambda_2$) to be associated with the even (odd) phase shift. Notice that wavelength dependence can be easily accounted for because the first condition must be satisfied at $\lambda_2$ only and the second one at $\lambda_1$ only. Solving for $L_A$ and $L_B$ we get

$$\begin{cases} L_A = \left[\dfrac{1-t+2m}{\kappa_1} + \dfrac{t+2k}{\kappa_2}\right]\dfrac{\pi}{16} - \dfrac{\delta L_2 + 3\delta L_1}{4} \\ L_B = \left[\dfrac{1-t+2m}{\kappa_1} - \dfrac{t+2k}{\kappa_2}\right]\dfrac{\pi}{8} + \dfrac{\delta L_2 - 3\delta L_1}{2} \end{cases}. \qquad (6)$$

In this case it is not straightforward to determine the selection rules for $k$ and $m$, because they will depend on the difference $\kappa_2 - \kappa_1$. Clearly the smaller this difference is, the better the rules in (4) apply also to this case. Once $\kappa(\lambda)$ and $\delta L(\lambda)$ are known, this simple formulas allow to design the directional couplers for doubly-point-symmetrical flattened filters. Furthermore, many different structures can be found with different choices of $t$, $k$, and $m$.

## *Numerical Example*

To confirm our analytical results we have performed some numerical simulation. We have considered silica buried waveguides, with a 4.5% index contrast and a 2μm x 2μm square cross section, which guarantees monomodality in the band of interest. We have calculated the propagation constants versus wavelength for the single waveguide mode and for the coupler supermodes using a fully vectorial commercial mode solver [10]. The inner wall waveguide separation of the couplers have been chosen to be 2μm. We have also checked with a commercial beam propagator [11] that, with this high index contrast, the contribution $\delta L(\lambda)$ is not only very small, but its wavelength dependence is the same of an equivalent straight coupler extra-length giving the same coupling. All these information can be implemented in a very accurate and simple transfer matrix model. In Fig 4 it is shown the spectral response for a 1490 nm/1550 nm splitter with the choice $t=1$, $m=1$, $k=0$. Notice that, by construction, to the wavelengths 1490 nm and 1550 nm correspond zeros in the port where they must be suppressed. Clearly, a fine tuning of the analytical solution, may give better performance over the whole channel band. Fig. 5 corresponds to the case $t=0$, $m=0$, $k=0$. In this case our receipt works only locally at the nominal wavelength, but does not guarantee flatness about 1550 nm. This can be qualitatively understood on the Poincaré sphere in Fig. 6. When the wavelength is not exactly 1550 nm the phase shift it is not an exact multiple of $\pi$. This give rise to a non-zero rotation about the $S_1$ axis corresponding to each phase shifter. This means a departure of the trajectory from the $S_1 S_3$ plane. Clearly, having chosen the doubly point-symmetric configuration, the sequence of phase shifts will be, for example, *ACAC*, where *A* (*C*) means Anticlockwise (Clockwise). But a *C* rotation can partially compensate for the effect of an *A* rotation if and only if they are performed starting from points on different hemispheres. On the contrary, when the points are in the same hemisphere a *C* rotation worsen the effect of the previous *A* rotation (in Fig.6 both rotation move the trajectory far away from the $S_1 S_3$ plane), and vice versa, giving a strong departure from the ideal trajectory, that is a small tolerance to wavelength changes. In this case the right choice is to correct the *A* rotation with another *A* rotation as shown in Fig.7. This means that we have to choose a singly point-symmetrical configuration *AACC*. In Fig. 8 it is shown the spectral response of this configuration, which is found to be well flattened on both ports. So when $2(\phi_i^B - \phi_i^A)$ and $2\phi_i^A$ belong to the same hemisphere, flatness may be guaranteed by the *AACC* scheme. But this is not always true as, for example, when choosing $t=0$, $m=1$, $k=-1$.

We have found a general rule to choose the symmetry that flattens both ports: the *AACC* configuration must be chosen if and only if

1) $2(\phi_i^B - \phi_i^A)$ and $2\phi_i^A$ belong to the same hemisphere AND $2(\phi_i^B + \phi_i^A)$ and $2\phi_i^A$ do not belong to adjacent quadrants (of the circle in the $S_1 S_3$ plane),

OR

2) $2(\phi_i^B + \phi_i^A)$ and $2\phi_i^A$ belong to opposite hemispheres AND $2(\phi_i^B - \phi_i^A)$ and $2\phi_i^A$ do not belong to opposite quadrants (e.g. when $t = 1$, $m = 2$, $k = 0$).

Notice that "do not belong to adjacent quadrants" is equivalent to "belong to opposite quadrants or to the same quadrant", as well as "do not belong to opposite quadrants" means "belong to adjacent quadrants or to the same quadrant".

## Conclusions

We have presented a geometrical and analytical approach to design interferometric band splitters. With this novel method we have generalized the doubly point symmetrical scheme for filter synthesis to the case of splitters made of wavelength dependent couplers. First a geometric interpretation of the working principle of these structures allows to derive simple analytical formulas for the case of wavelength independent couplers. Then the same formulas can be easily extended to the case of wavelength dependent couplers. Also the geometrical and physical insight helps us to broaden the class of flattened splitters to singly point symmetric structures, which are needed for certain combinations of coupling angles. The proposed method can be easily extended to any kind of interferometric filter featuring more couplers and phase shifters and/or different symmetries.

**Figure captions**

Fig. 1. The doubly point symmetric structure. First the building block, composed of a type A coupler and a half type B coupler, is repeated point-symmetrically, resulting in a ABA structure. This structure is repeated point-symmetrically again to give the desired result.

Fig. 2. Generalized Poincaré sphere for the analysis of two coupled waveguides. The generic point P is represented together with its relative phase angle $\theta$ and power splitting angle $\alpha$. Physical transformation are represented by composition of rotations about axes on the $S_1S_2$ plane. The points on the rotation axis represent the eigenstates of the system. In particular a synchronous coupler with coupling angle $\phi$ is represented by a $2\phi$ rotation about the $S_2$ axis, whereas a $\theta$ phase shift is represented by a $\theta$ rotation about the $S_1$ axis.

Fig. 3. Geometric representation on the $S_1S_3$ plane of the action of a Mach Zehnder interferometer in the two cases of (a) even and (b) odd phase shifts. A coupler with coupling angle $\phi$ gives rise to a $2\phi$ rotation, a $2n\pi$ phase shift lives the system as it is whereas a $(2n+1)\pi$ phase shift turns over the state with respect to the $S_1$ axis.

Fig. 4. Simulated spectral response for the bar port (continuous line) and the cross port (dashed line) of a splitter with $t=1, m=1, k=0$, ideally designed for separating a 1490 nm channel in the bar port from a 1550 nm channel in the cross port. The doubly point symmetric structure ensures flatness for both ports.

Fig. 5. Simulated spectral response for the bar port (continuous line) and the cross port (dashed line) of a splitter with $t=0, m=0, k=0$, ideally designed for separating a 1490 nm channel in the cross port from a 1550 nm channel in the bar port. In this case the response is not flat for the bar port.

Fig. 6. Geometrical representation on the full 3D Poincaré sphere of the response in Fig. 5 for a wavelength slightly different by 1550 nm, corresponding to phase shifts slightly greater than π. The sequence *ACAC* moves the trajectory on the spherical surface away from the $S_1S_3$ plane, since the angular excess ad up at each stage. This means that, unlike the nominal wavelength case (which trajectory remains confined in the $S_1S_3$ plane), the coupler rotations are performed at increasing distance from the $S_1S_3$ plane, that is far from the nominal ending point $E_1$.

Fig.7. Same as in Fig. 6 when the sequence *AACC* is chosen. In this case the trajectory remains close to the $S_1S_3$ plane. The angular excess of the second (fourth) phase shifter compensates the angular excess of the first (third) phase shifter. In this way all the coupler rotations are performed close to the $S_1S_3$ plane and they add up to almost zero, like at the nominal wavelength.

Fig. 8. Spectral response of a $t = 0$, $m = 0$, $k = 0$ structure when the *AACC* configuration is chosen. In this case, as explained in Fig 7, the 1550 nm port is very well flattened.

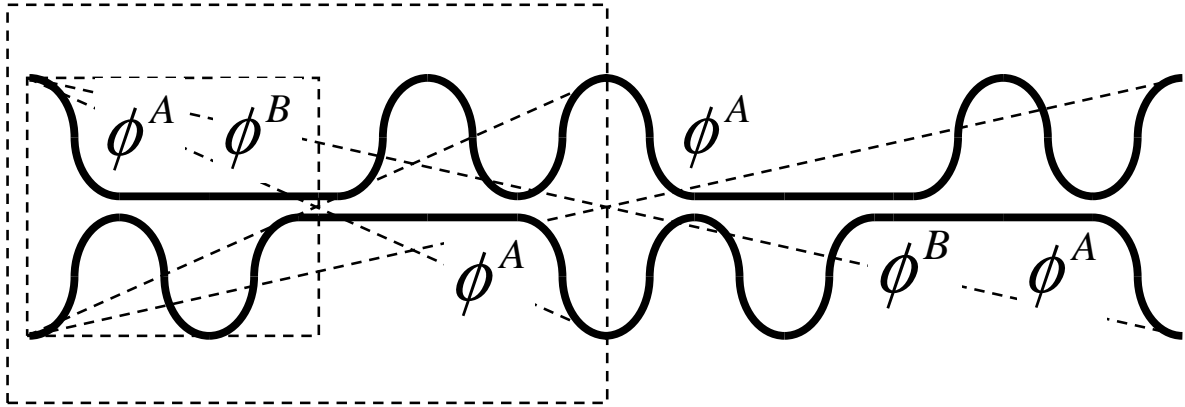

Fig. 1

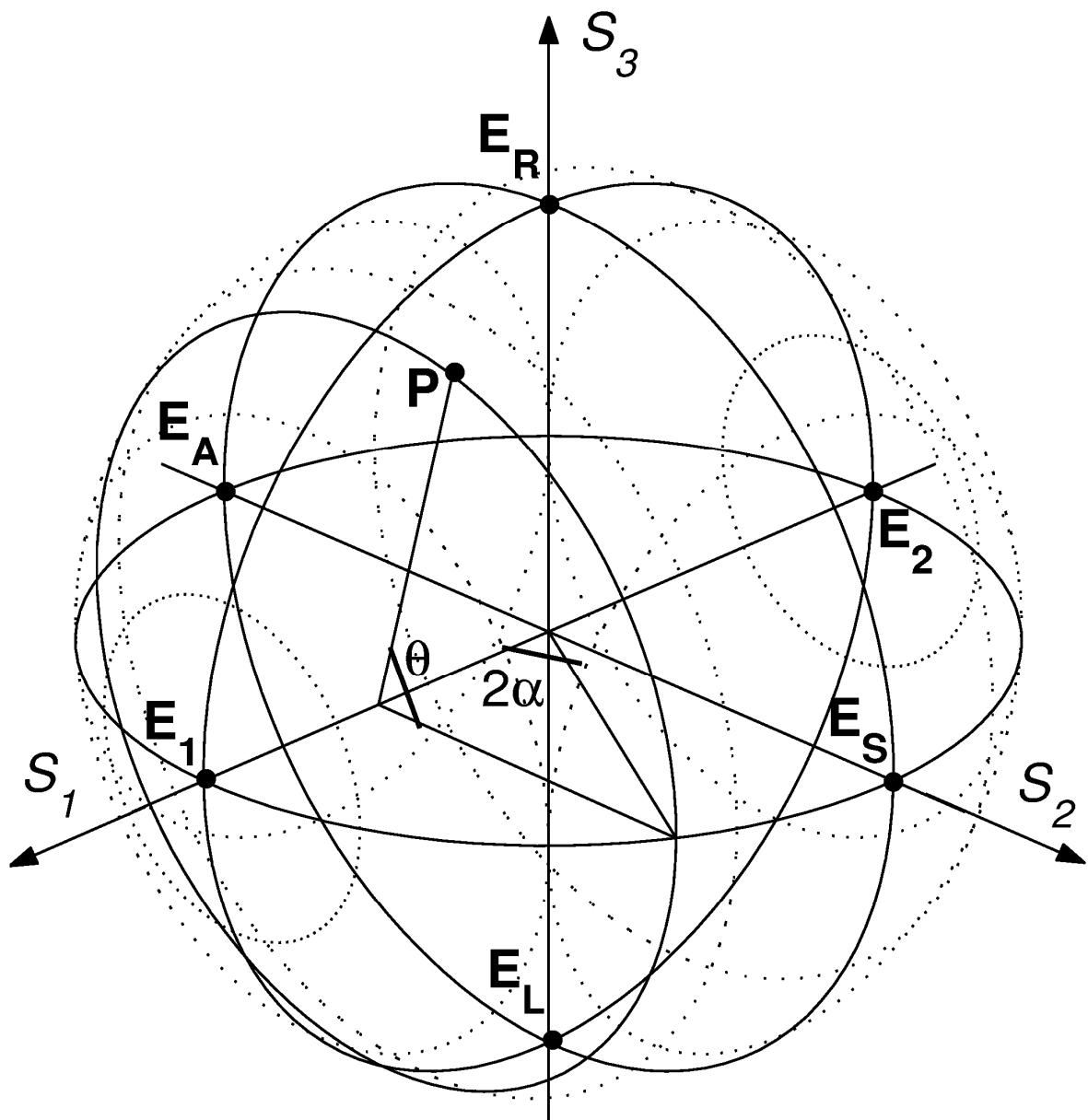

Fig. 2

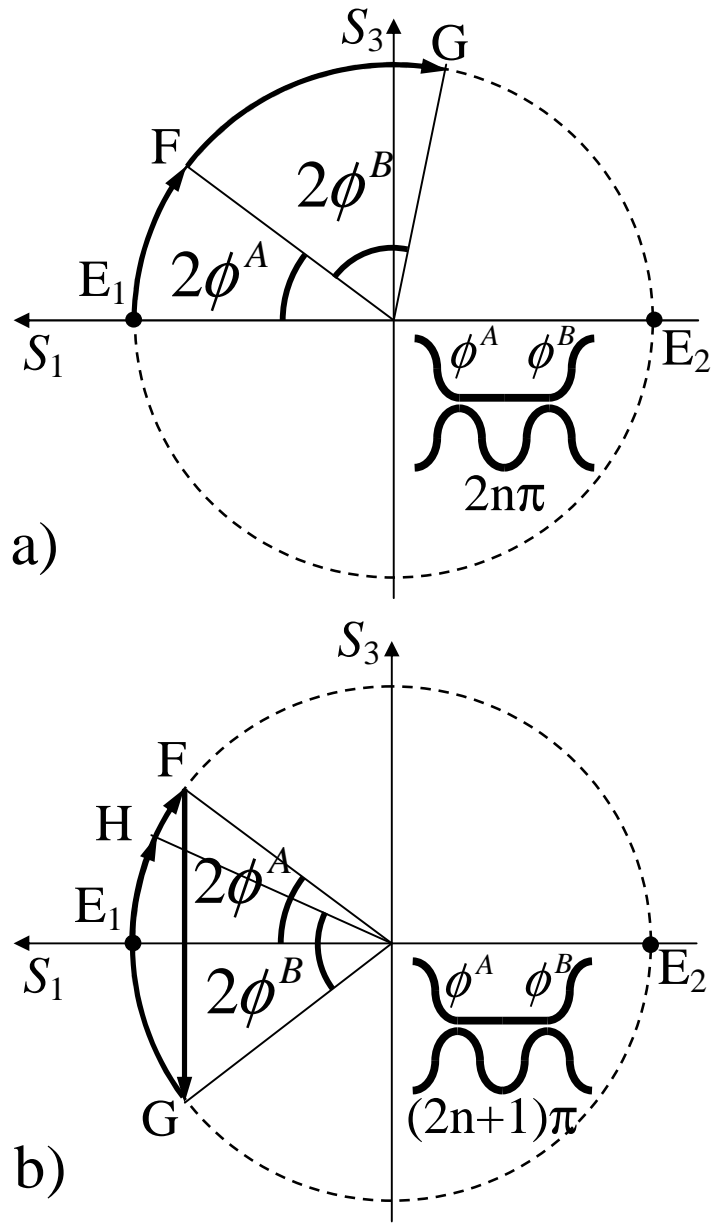

Fig. 3

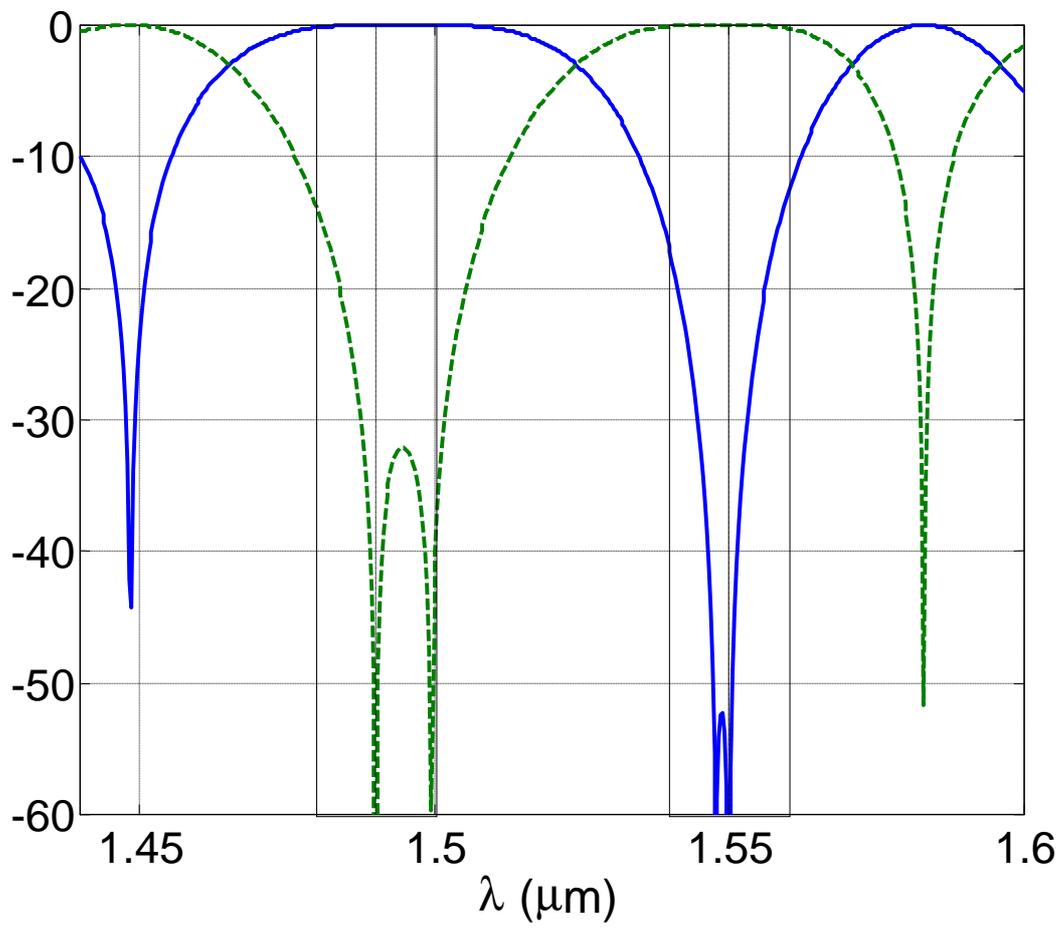

Fig. 4

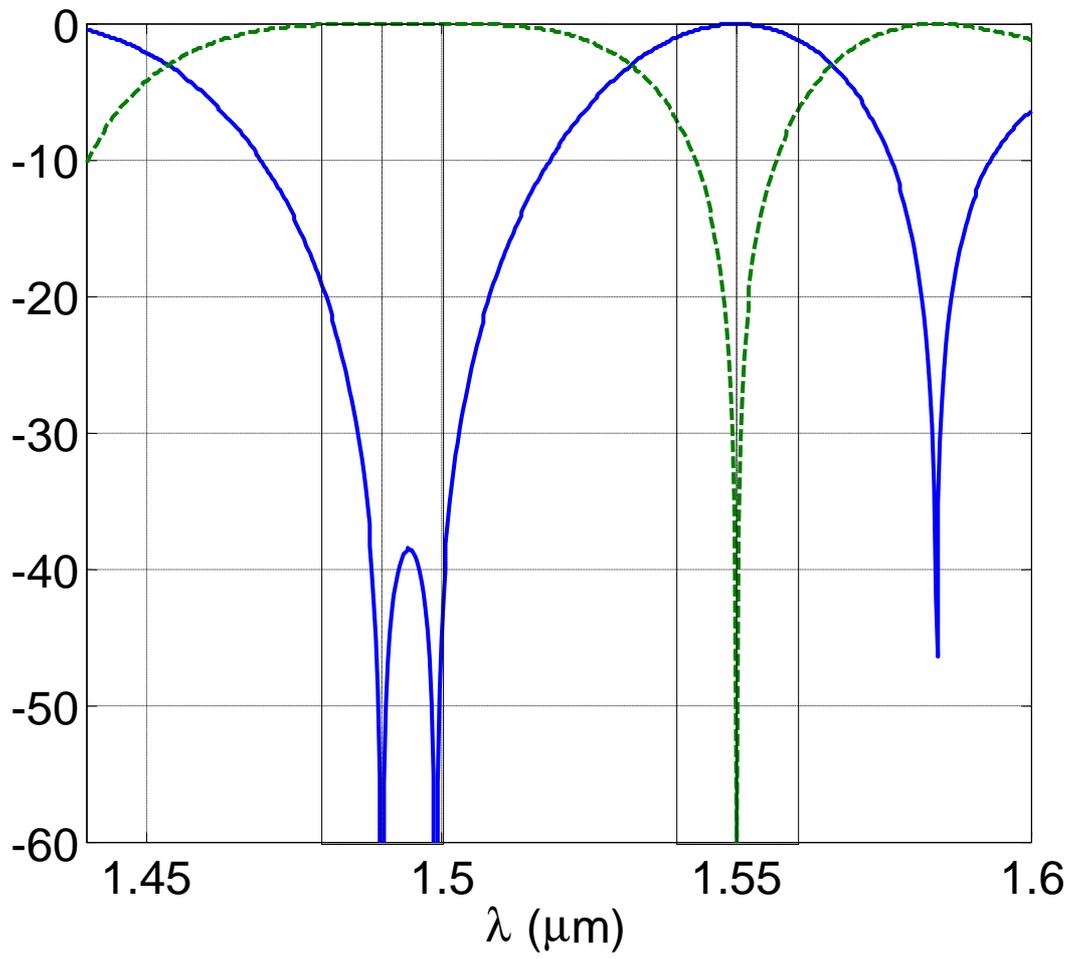

Fig. 5.

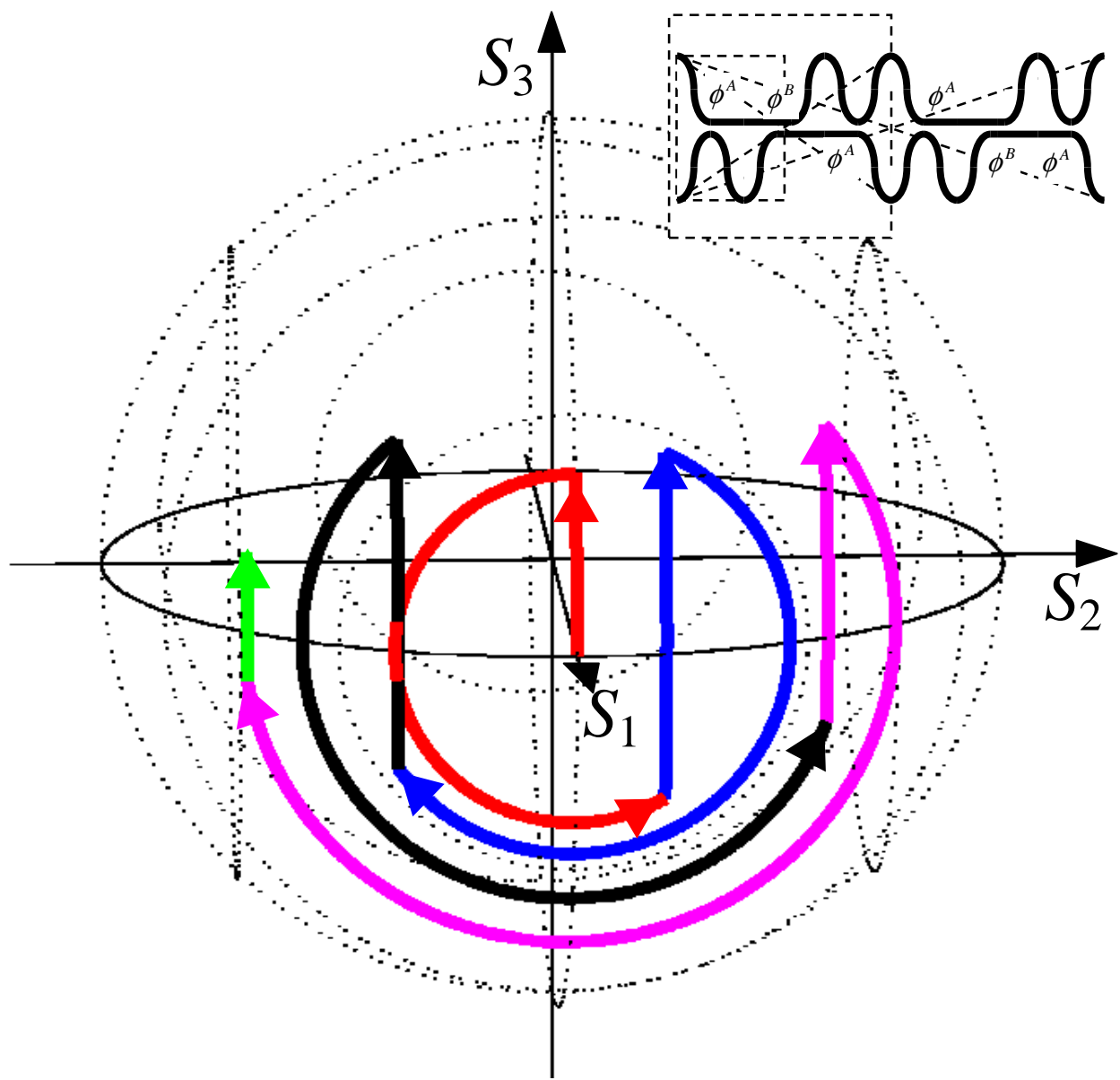

Fig. 6

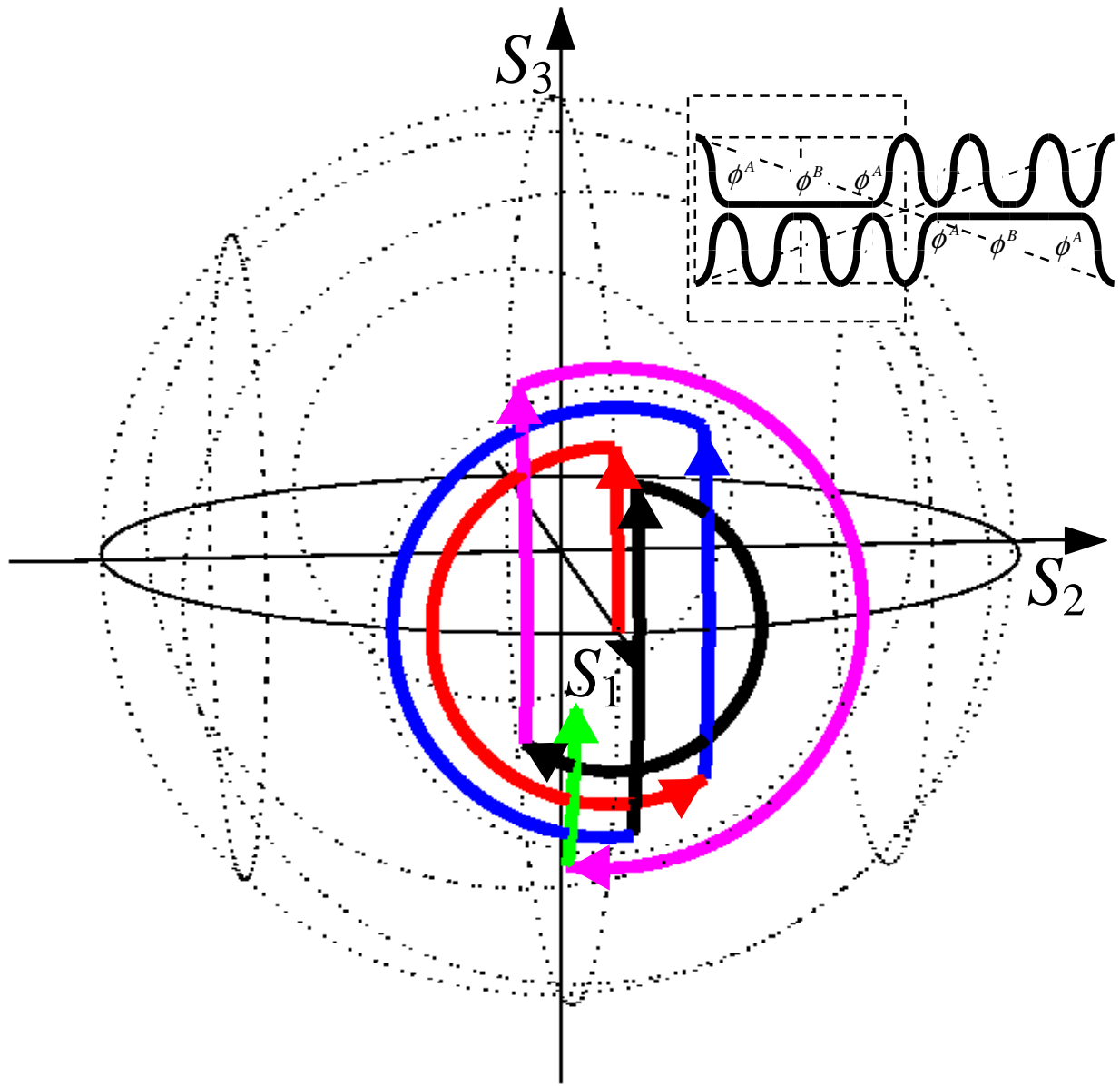

Fig. 7

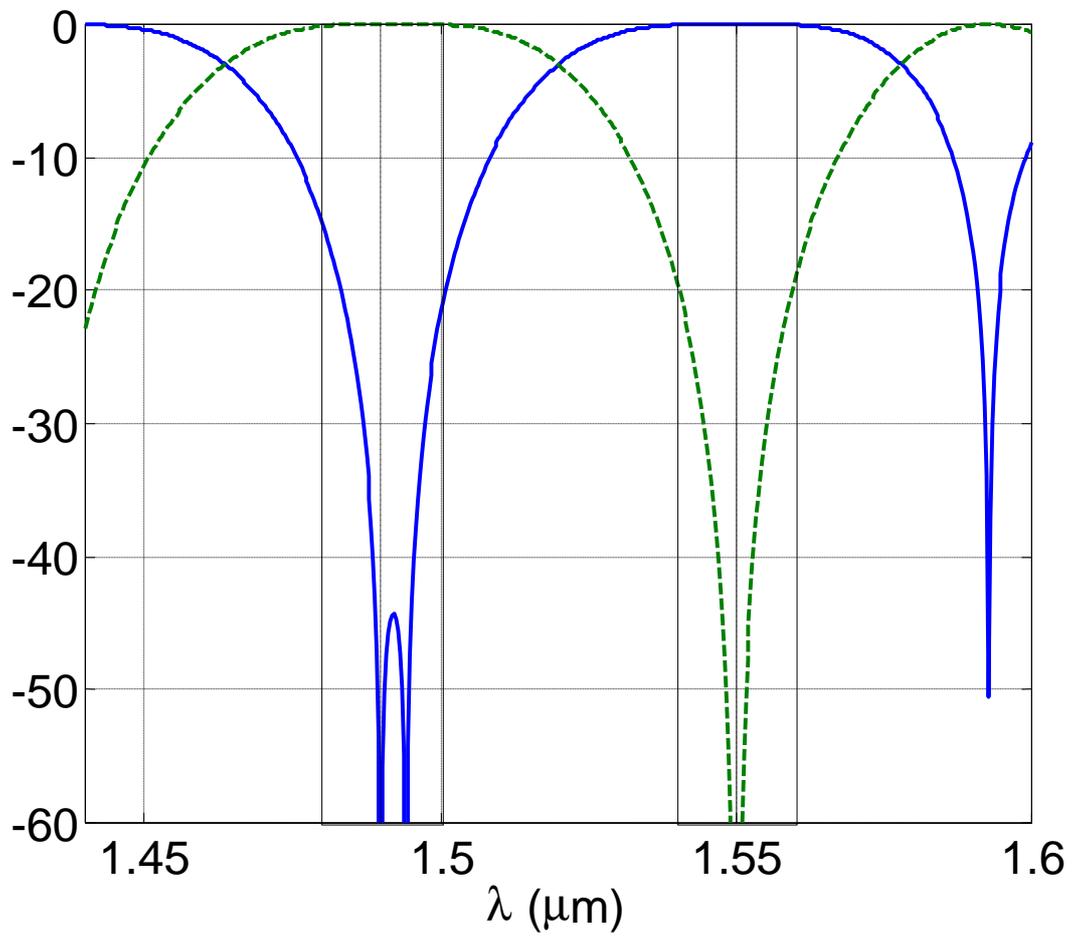

Fig. 8